\documentclass[11pt,a4paper]{article}
\usepackage{amsmath,amssymb}
\usepackage{epsfig,graphicx}

\usepackage{amsmath,amssymb}
\usepackage{epsfig,graphicx}

\begin{document}

\title{\bf Spontaneous radiatively induced breaking of conformal invariance
in the Standard Model}
\author{A.~B.~Arbuzov$^{a,b}$\footnote{Based on the talk presented by A.B.~Arbuzov at QUARKS-2014, Suzdal, Russia.},
V.~N.~Pervushin$^a$, R.~G.~Nazmitdinov$^{a,c}$, A.~E.~Pavlov$^d$, A.~F.~Zakharov$^{a,e}$
\\
$^a$ \small{\em Bogoliubov Laboratory of Theoretical Physics,
Joint Institute for Nuclear Research, Dubna, 141980 Russia} \\
$^b$ \small{\em Department of Higher Mathematics, University Dubna,
Dubna, 141980, Russia} \\
$^c$ \small{\em Department de F{\'\i}sica, Universitat de les Illes Balears,
Palma de Mallorca, E-07122, Spain} \\
$^d$ \small{\em Moscow State Agri-Engineering University, Moscow, 127550, Russia}\\
$^e$ \small{\em Institute of Theoretical and Experimental Physics, Moscow, 117259, Russia}
}

\date{}
\maketitle

\begin{abstract}
A mechanism of radiatively induced breaking of the conformal symmetry
in the Standard Model is suggested. The system of one scalar (Higgs) and one
fermion (top-quark) fields with Yukawa and $\phi^4$ interactions is considered.
The infrared instability of the Coleman-Weinberg effective potential for this system
leads to the appearance of a finite renormalization scale and thus to
breaking of the conformal symmetry. Finite condensates of both scalar and spinor
fields appear. It is shown that the top quark condensate can
supersede the tachyon mass of the Higgs field.
The Higgs boson is treated as an elementary scalar and the standard mechanism of
electroweak symmetry breaking remains unchanged.
The difference from the Standard Model appears in the value of the Higgs boson
self-coupling constant.
\end{abstract}

\section{Introduction}

Recently the major LHC experiments reported upon the discovery of a boson
with the mass of about $125$~GeV \cite{:2012gk,:2012gu}.
Further experimental studies~\cite{Aad:2013xqa,Chatrchyan:2013lba} showed
that this particle behaves very much like the Higgs boson of
the Standard Model (SM)~\cite{SM}. Nevertheless, the question about the
fundamental mechanism(s) of mass generation is far beyond the final
resolution. Moreover, there is a number of indirect evidences that the
conformal symmetry (CS) might be the proper feature of the true fundamental
theory, while the SM is just an effective theory with a softly broken
CS, see {\it e.g.}~\cite{Gorsky:2014una} and references therein.

According to the general wisdom, all SM particles (may be except neutrinos) own
masses due to ``interaction'' with the Higgs boson vacuum expectation value. 
The latter emerges after the spontaneous breaking of the $O(4)$ symmetry in the scalar
sector~\cite{Englert:1964et,Higgs:1964pj}.
In the SM, one deals with the potential
 \begin{equation}  \label{5a}
 V_{\rm Higgs}(\Phi)=\lambda(\Phi^\dagger\Phi)^2 + \mu^2\Phi^\dagger\Phi,
 \end{equation}
where one component of the complex scalar doublet field
$\Phi=\left(\begin{array}{c}\Phi^+\\ \Phi^0\end{array}\right)$
acquires a non-zero vacuum expectation value $\langle\Phi^0 \rangle = v/\sqrt{2}$
if $\mu^2<0$ (the vacuum stability condition $\lambda>0$ is assumed).
Note that the tachyon-like mass term in the potential is crucial for this construction.
In contrast to the $O(4)$ spontaneous symmetry breaking (SSB),
it breaks the conformal symmetry explicitly being
the only one {\em fundamental} dimensionful parameter in the SM.
We recall that the explicit conformal symmetry breaking in the
Higgs sector gives rise to the naturalness (or fine tuning) problem in the renormalization
of the Higgs boson mass.
That is certainly the most unpleasant feature of the SM.

\section{The naturalness problem}

Let us look at some details of the naturalness problem. In the one-loop
approximation the Higgs boson mass gets huge corrections due to
quadratically divergent amplitudes:
\begin{eqnarray} \label{naturalness}
M_H^2 = (M_H^0)^2 + \frac{3\Lambda^2}{8\pi^2v^2}\biggl[M_H^2 + 2M_W^2 + M_Z^2 - 4m_t^2 \biggr],
\end{eqnarray}
where $\Lambda$ is an ultraviolet cut-off parameter.
Certainly it is unnatural to have a huge hierarchy between $M_H$
and $M_H^0$\footnote{We stress that $M_H^0$ here should be directly related to the tachyon-like
mass parameter in the initial Lagrangian, where it appears as a {\it fundamental} scale.}. 
There are two general ways to solve the problem: \\
--- either to exploit some (super)symmetry to cancel out the huge terms, \\
--- or to introduce some new physics at a scale not very far from the
electroweak (EW) one, {\it i.e.} making $\Lambda$ being not large.

One can find in the literature quite a lot of models for both options.
Actually, since the (super)symmetry is not observed experimentally at the EW scale,
the first way besides the introduction of the symmetry requires application
of a mechanism of its breaking at some energy scale close to the EW one. 
On the other hand, the experimental data coming from modern accelerators and rare decay
studies disfavors most of scenarios of new physics with scales up to about 1~TeV
and even higher.
Moreover, it was shown that the measured value of the Higgs boson mass makes the SM
being self-consistent up to very high energies of the order
$10^{11}$~GeV~\cite{Bednyakov:2012rb} or even
up to the Planck mass scale~\cite{Bezrukov:2012sa,Alekhin:2012py}.
Direct and indirect experimental searches push up and up possible energy scale of new physical
phenomena.
So the naturalness problem becomes nowadays more and more prominent.
And the question, why the top quark mass, the Higgs boson mass, the vacuum 
expectation value $v$, and the electroweak scale itself 
are of the same order becomes more and more intriguing.

The correction (\ref{naturalness})
comes from Feynman diagrams with boson single-propagator loops (tadpoles) and from
the two-point function with two top-quark propagators. The latter actually
is reduced to a top quark tadpole:
\begin{eqnarray}
&& -N_c\int_{\Lambda_t}\frac{d^4k}{i\pi^2}\,
\frac{\mathrm{Tr}(\hat{k}+m_t)((\hat{p}-\hat{k})+m_t)}{(k^2-m_t^2)((p-k)^2-m_t^2}
\to
-4N_c\int_{\Lambda_t}\frac{d^4k}{i\pi^2}\,\frac{1}{k^2-m_t^2} + \mathcal{O}(m_t^2)
 \nonumber \\
&& \quad = -4N_cA_0(m_t^2,\Lambda_t^2) +  \mathcal{O}(m_t^2),
\end{eqnarray}
where $A_0$ is the standard Passarino-Veltman function.

One the other hand, we have the following standard formal definition of
the quark condensate:
\begin{eqnarray}
\langle \bar{q}\, q\rangle \equiv
-N_C\int_{\Lambda_q}\frac{d^4k}{i\pi^2}\,\frac{\mathrm{Tr}(\hat{k}+m_q)}{k^2-m_q^2+i\varepsilon}
\sim -4N_Cm_q A_0(m_q^2,\Lambda_q^2).
\end{eqnarray}

So, the top quark contribution to Eq.~(\ref{naturalness}) is {\em formally} provided
by its condensate value. Our conjecture is that the formal correspondence has a deep
physical meaning and reveals itself not only here.

We claim that the very existence of the quark condensate comes out from
the QFT rules. But its value of course depends on the details of the model. In particular,
the value of the light quark condensate is rather well known from
low-energy strong interactions, $\sqrt[3]{\langle \bar{q}\, q\rangle} \simeq - 250$~MeV.
The possibility to extract this number from observables if provided by the
presence of [non-trivial] non-perturbative interactions at the corresponding energy
scale. On the other hand, nothing at all is known from the phenomenology
concerning the value of the top quark condensate. That is just because the
energy scale of the top quark mass allows only perturbative QCD interactions,
which are not sensitive to the condensate\footnote{Due to the Furry theorem the coefficient
before the fermion tadpole with vector vertex is zero. While the tadpole itself
(with a scalar vertex) can be non-zero.}. Note also, that it is clear that the scale of
the light quark condensate is provided by the $\Lambda_{\mathrm{QCD}}$ scale:
$\langle \bar{q}\, q\rangle \sim - M_q\Lambda_{\mathrm{QCD}}^2$, where $M_q$
is the constituent quark mass which in its turn also has the same scale
$M_q \sim \Lambda_{\mathrm{QCD}}$.

\section{Coleman-Weinberg effective potential in the SM}

Let us now consider a simple model with one scalar field $\phi$
and one fermion field $f$.
We demand the conformal symmetry for the model. The symmetry
allows\footnote{For terms in the Lagrangian of a renormalizable QFT model
{\em allowed} is practically equivalent to {\em must have}.}
the existence of two types of interactions in this models:
the $\phi^4$ self-interaction of the scalar and the Yukawa term.
So we start with the classical potential
\begin{eqnarray} \label{V_CW_cl}
V_{\mathrm{cl}}=\lambda\phi_c^4/4! + y\phi_c\bar{f}_c f_c,
\end{eqnarray}
where we used the notation of Ref.~\cite{Coleman:1973jx}, in particular
the subscript ``c'' underlines that $\phi_c$ and $f_c$ are classical fields and they
obey the conformal symmetry. The standard one-loop calculation of the
effective potential gives two contributions: the one from scalar loops,
and the one from fermion loops:
\begin{eqnarray} \label{delta_V_sc}
&& \Delta V_{\mathrm{sc}}= \frac{1}{2}\int\frac{d^4k}{(2\pi^4)}\ln\left(
1 + \frac{\lambda\phi_c^2}{2k^2}\right) \to \frac{\lambda\Lambda^2}{256\pi^2}\phi_c^2
+ \frac{\lambda^2\phi_c^4}{256\pi^2}\left(\ln\frac{\lambda\phi_c^2}{2\Lambda^2}
-\frac{1}{2}\right),
\\ \nonumber
&& \Delta V_{f} = -4N_C{\mathrm Tr}\int\frac{d^4k}{(2\pi^4)}\ln\left(
1 + \frac{y\phi_c(\hat{k}+m_f)}{k^2-m_f^2}\right) \to
-4N_C\frac{ym_f\Lambda^2}{16\pi^2}\phi_c
\\ \label{delta_V_f}
&& \qquad -4N_C\frac{y^2m_f^2\phi_c^2}{32\pi^2}\left(\ln\frac{ym_f\phi_c}{\Lambda^2}
-\frac{1}{2}\right)+\ldots
\end{eqnarray}
Due to the condition of the classical conformal symmetry, we have to remormalize the
first term on the right hand side of Eq.~(\ref{delta_V_sc}) to zero. On the other
hand, it is clear that the effective potential possesses infrared divergence at $\phi=0$.
So, some energy scale $M$ should be introduced to renormalize the logarithmic term.
As demonstrated by S.~Coleman and E.~Weinberg, that induces a spontaneous breaking
of the conformal symmetry and leads to the appearance of a non-zero mass and a non-zero
vacuum expectation value of $\phi$. Obviously, if we have
 $\langle\phi\rangle\neq 0$ in a model with Yukawa interactions, we automatically get
a mass for the fermion. Then the second contribution~(\ref{delta_V_f})
to the effective potential emerges. Note that the conformal symmetry doesn't
require to drop ({\it i.e.} renormalize to 0) the first term on the right hand side there,
since it is proportional to $m_f$ which is zero in the unbroken phase. This term,
is again nothing else but the tadpole contribution, {\it i.e.} the fermion condensate. In this way we
have two phases: the classical one with
$$ m_\phi=m_f\equiv 0,\qquad \langle \phi\rangle \equiv 0, \qquad
\langle \bar{f}\, f\rangle \equiv 0$$
and the one with spontaneously broken CS:
$$
m_\phi\sim m_f\sim M,\qquad
\langle \phi\rangle \sim M, \qquad
\langle \bar{f}\, f\rangle \sim -M^3,
$$
where $M$ is the renormalization scale, and we assumed that the coupling
constants are not extremely small $\lambda\sim y\sim 1$.

So, we clearly see that system~(\ref{V_CW_cl}) is unstable in the infrared region,
which leads to the effect of dimensional transmutation. According to the logic of
the original paper~\cite{Coleman:1973jx}, the scale comes to the model somewhere
from outside the theory via the renormalization procedure. Another crucial
point is the question about stability and perturbativity in the phase with the
broken conformal symmetry.

\section{Dimensional Transmutation in the SM}

We suggested~\cite{Pervushin:2012dt} the following minimal modification of
the Standard model: let us drop the tachyon mass term from the Lagrangian.
We take the most intensive interaction of the Higgs filed $h$
\begin{eqnarray}
L_{\mathrm{int}} = - \frac{\lambda}{4}h^4 - \frac{y_t}{\sqrt{2}}h\;\bar{t}t,
\end{eqnarray}
where $h$ is related to the initial field $\Phi$ in the standard way.
In this case we have a model with a classical conformal invariance.
We claim that the apparent breaking of this symmetry can happen spontaneously
because of the infrared instability at the quantum level. Obviously,
in such a case, see {\it e.g.} Ref.~\cite{Bardeen:1995kv}, the softly broken classical symmetry
will protect the Higgs boson mass from rapid running in the UV region.
So let us look at a stable solution in the broken phase. The leading contribution
to the Coleman-Weinberg effective potential comes from the top quark tadpole:
\begin{eqnarray}
V_{\rm eff}(h) \approx \frac{\lambda}{4}h^4
+ \frac{y_t}{\sqrt{2}}\langle \bar t\,t\rangle h.
\end{eqnarray}
Naturally we choose the electroweak energy scale as the (re)normalization point.
Than all dimensionful parameters in the effective potential are defined
by this scale.
The extremum condition for the potential $dV_{\rm cond}/dh|_{h=v}=0$
yields the relation
 \begin{equation} \label{lambda}
 \lambda {v^3} = -\frac{y_t}{\sqrt{2}} \langle \bar t\, t\rangle.
 \end{equation}
It follows from the fact that
the Higgs field has a zero harmonic $v$ in the standard decomposition
of the field $h$ over harmonics  $h = v  + H$, where $H$ represents excitations
(non-zero harmonics) with the condition $\int d^3x H=0$. The Yukawa coupling of
the top quark $y_t = 0.995(5)$ is known from the experimental value
of top quark mass $m_t=v y_t/\sqrt{2}\simeq 173.2(9)$~GeV~\cite{Agashe:2014kda},
and $v=(\sqrt{2}G_{\mathrm{Fermi}})^{-1/2}\approx 246.22$~GeV is
related to the Fermi coupling constant derived from the muon life time measurements.
So, the spontaneous symmetry breaking yields the potential minimum which results in the non-zero
vacuum expectation value $v$ and the Higgs boson mass.
In fact, the substitution $h = v  + H$  gives
\begin{equation}
V_{\rm eff}(h) = V_{\rm eff}(v) + \frac{m_H^2}{2}H^2
+ \lambda^2 v H^3+\frac{\lambda}{4}H^4,
\end{equation}
which defines the scalar particle mass as
\begin{equation}
\label{mh}
m_H^2= 3\lambda v^2.
\end{equation}
We stress that this relation is different from the one $(m_H^2=2\lambda v^2)$
which emerges in the SM with the standard Higgs potential~(\ref{5a}).

With the aid of Eqs.~(\ref{lambda}) and (\ref{mh}),
the squared scalar particle mass can be expressed in terms of the top
quark condensate:
\begin{equation}\label{h_mass}
m^2_H = -\frac{3 y_t \langle \bar t\, t\rangle}{\sqrt{2}v}.
\end{equation}
To get $m_H=125.7$~GeV we need
\begin{equation}
\langle \bar t\, t\rangle \approx -(122\ {\mathrm{GeV}})^3.
\end{equation}
As discussed above, such a large value of the top quark condensate does not
affect the low energy QCD phenomenology. Since heavy quark condensates do not
contribute to observed QCD quantites ({\it e.g.} via sum rules), we do not have
any experimental or theoretical limits on the top condensate value, 
see~\cite{Cvetic:1997eb} and references therein.

Note that the energy scale of the
top quark condensate appears to be the same as the general electroweak one.
We believe that the scale of light quark condensate
is related to the scale of the conformal anomaly in QCD. At the same time
those anomalous properties of the QCD vacuum lead to the constituent
mass of a light quark to be of the order $300$~MeV. As concerning the top
quark, some anomalous properties of the {\em relevant} vacuum
give rise to the mass of this quark\footnote{Certainly, QCD effects both
in the mass and in the condensate value of the top quark are small
compared to the Yukawa ones.} and to the condensate being of the same energy scale.

One can note that even we have dropped the scalar field mass term from the
classical Lagrangian, it will re-appear after quantization and subsequent renormalization.
In fact, such a counter-term in the Higgs sector is necessary.
But as described in Ref.~\cite{Bardeen:1995kv}, the conformal symmetry of the classical
Lagrangian will lead to just the proper quantity in the mass term being consistent
with all other quantum effects.
A similar situation takes place in QCD: the chiral symmetry at the quark level
re-appear at the hadronic level even so that the breaking is obvious~\cite{Witten:1983tw}.
  
In conclusion, we suggest to apply the Coleman-Weinberg mechanism of dimensional transmutation
to induce spontaneous conformal symmetry breaking in the Standard Model.
This enables us to avoid the problem of the regularization of the divergent
tadpole loop integrals by relating them to condensate values hopefully extracted from
experimental observations.
The top quark condensate can supersede the tachyon-like mass term in the Higgs potential.
The suggested mechanism allows to establish relations between condensates and masses including
the Higgs boson one.
In a sense, we suggest a simple bootstrap between the Higgs and top
fields (and their condensates).
We underline that we consider the Higgs boson to be an elementary particle
without introduction of any additional interaction beyond the SM contrary to
various technicolor models.
After the spontaneous symmetry breaking in the tree level
Lagrangian, the difference from the SM appears only in the value of the
Higgs boson self-coupling $\lambda$. The latter hardly can be extracted from the LHC
data, but it will be certainly measured at a future linear $e^+e^-$ collider.

{\bf Acknowledgments} A.Arbuzov is grateful to the Dynasty foundation for financial support.

\end{document}